\documentclass[namedreferences]{solarphysics}

\usepackage[hyperref,optionalrh]{spr-sola-addons} 
\usepackage[inline]{enumitem}
\usepackage{graphicx}        
\usepackage{color}           
\usepackage{breakurl}        

\usepackage[normalem]{ulem}

\newcommand {\pry}[1]{#1}

\ifx\urlurl    \undefined
\def\urlurl#1{\href{http://#1}{\textsf{#1}}}\fi

\newcommand{\ion}[2]{{#1} {\sc #2}}

\chardef\us=`\_

\begin{document}

\begin{article}
\begin{opening}

\title{An Analysis of Spikes in \emph{Atmospheric Imaging Assembly} (AIA) Data}

\author[addressref={aff1,aff2},corref,email={peter.r.young@nasa.gov}]{\inits{P.R.}\fnm{Peter R.}~\lnm{Young}\orcid{0000-0001-9034-2925}}
\author[addressref={aff1}]{\inits{N.M.}\fnm{Nicholeen M.}~\lnm{Viall}\orcid{0000-0003-1692-1704}}
\author[addressref={aff1,aff3}]{\inits{M.S.}\fnm{Michael S.}~\lnm{Kirk}\orcid{0000-0001-9874-1429}}
\author[addressref={aff1}]{\inits{E.I.}\fnm{Emily I.}~\lnm{Mason}\orcid{0000-0002-8767-7182}}
\author[addressref={aff4}]{\inits{L.P.}\fnm{Lakshmi Pradeep}~\lnm{Chitta}\orcid{0000-0002-9270-6785}}

\address[id=aff1]{NASA Goddard Space Flight Center, Code 671, Heliophysics Science Division, Greenbelt, MD 20771, USA}
\address[id=aff2]{Department of Mathematics, Physics and Electrical Engineering, Northumbria University, Newcastle upon Tyne, UK}
\address[id=aff3]{ASTRA, LLC, 282 Century Place, Suite 1000, Louisville, Colorado, 80027, USA}
\address[id=aff4]{Max Planck Institute for Solar System Research, Justus-von-Liebig-Weg 3, D-37077 Goettingen, Germany}

\runningauthor{P.R.~Young et al.}
\runningtitle{AIA Spikes}

\begin{abstract}
The \emph{Atmospheric Imaging Assembly} (AIA) onboard the \emph{Solar Dynamics Observatory} (SDO) returns high-resolution images of the solar atmosphere in seven extreme ultraviolet (EUV) wavelength channels. The images are processed on the ground to remove intensity spikes arising from energetic particles hitting the instrument, and the despiked images are provided to the community.
 In this article a three-hour series of images from the 171\,\AA\ channel obtained on 28 February 2017 was studied to investigate how often the despiking algorithm  gave false positives caused by compact brightenings in the solar atmosphere. The latter were identified through spikes appearing in the same detector pixel for three consecutive frames. 1096 examples were found from the 900 image frames. These ``three-spikes" were assigned to 126 dynamic solar features, and it is estimated that the three-spike method identifies 20\,\%\ of the total number of features affected by despiking. For any ten-minute sequence of AIA 171\,\AA\ images there are therefore around 35 solar features that have their intensity modified by despiking. The features are found in active regions, quiet Sun, and coronal holes and, in relation to solar surface area, there is a greater proportion within coronal holes. In 96\,\%\ of the cases, the despiked structure is a compact brightening of size two arcsec or less and the remaining 4\,\%\ have narrow, elongated structures. \pry{By applying an EUV burst detection algorithm, we found that  96\,\%\ of the events could be classed as EUV bursts.}
\pry{None of the spike events are} rendered invisible by the AIA processing pipeline, but the total intensity over an event's lifetime can be reduced by up to 67\,\%.  Users are recommended to always restore the original intensities to AIA data when studying short-lived or rapidly evolving features that exhibit fine-scale structure.
\end{abstract}
\end{opening}

\section{Introduction}
     \label{Sectionintro} 

Most datasets from instruments that observe the Sun exhibit artifacts that hamper interpretation of the features under study. The artifacts are often internal to the instrument, such as dust or particle contamination on optical surfaces and fixed pattern noise or hot pixels on the detector. An important external artifact for some space-borne instruments arises from energetic particles hitting silicon detectors, generating signals that mimic those from the solar photons being observed. The particles can be from the Earth's radiation belts, solar energetic particles, or galactic cosmic rays. It is standard practice for the instruments' calibration pipelines to remove the various artifacts, giving the users cleaned data-sets appropriate for scientific analysis. In some cases, however, the solar data  can be adversely affected by these treatments, potentially impacting interpretation or measurements. The present article concerns extreme ultraviolet images from the \textit{Atmospheric Imaging Assembly} \citep[AIA:][]{2012SoPh..275...17L} onboard the \textit{Solar Dynamics Observatory} \citep[SDO:][]{2012SoPh..275....3P}. In particular, it addresses how often imaged events in the solar atmosphere are affected by the algorithm that removes energetic-particle hits from the images.

This issue is important because it is the despiked images that are distributed to the scientific community, not the original images. Thus scientists may be unaware that signal has been removed from the features that they are studying. The AIA team does provide an auxiliary ``spikes" file for each image that stores the location and intensity of each spike, and so it is possible to reconstruct the original image. The  spikes files are of interest in their own right. They are factors of 100 to 200 smaller than the AIA image files and so they are more amenable to large-scale processing than the images, given that AIA has returned around \pry{two hundred million} images since 2010.  If we can separate the spikes that belong to solar features from those due to energetic-particle hits, then we have a ready-made database of compact, dynamic solar features that extends over the entire SDO mission. The present article provides a method that is able to separate solar features from energetic-particle hits using only the information in the spikes files.

For AIA data, the energetic particles typically only have a significant impact on one pixel, giving rise to an anomalous intensity spike in the image, although  with weak residual signal on directly adjacent pixels. If the particle's path is at a shallow angle to the detector surface, then it can also yield a streak extending over multiple pixels, although these are much less common. For AIA, the particles can also release high-energy photons after impacting the instrument structure which may then also lead to image spikes \pry{\citep{2012SoPh..275...17L}}.

For this article we use the term \emph{spike} to refer to a single pixel that has been flagged by the AIA calibration-pipeline despiking algorithm. A \emph{real} or \emph{true} spike is one that corresponds to a solar atmospheric structure rather than a particle hit. A spike \emph{event} is a group of one or more true spikes that occur \pry{close together in space and time} as part of the evolution of a solar atmospheric structure.

\citet{2013ApJ...766..127Y} analyzed AIA data of a set of compact brightenings along a flare ribbon observed on 16 February 2011 and found that some of the pixels' intensities had been erroneously adjusted by the AIA despiking algorithm.  A similar problem was found in studies of coronal-hole jets by \citet{2014PASJ...66S..12Y,2014SoPh..289.3313Y}, suggesting that any feature showing intense, compact brightenings could suffer from this effect. 

\pry{\emph{The Guide to SDO Data Analysis} \citep{sdo_guide} states clearly in Section~7.3 that the AIA despiking algorithm may remove real solar features. However, a search of the literature suggests that it is relatively rare for users to respike AIA data as part of their analysis.} The full text of articles in the journals \emph{Solar Physics}, \emph{The Astrophysical Journal}, and \emph{Astronomy \& Astrophysics} were searched for terms such as ``respike", ``re-spike", and variants thereof, and only five were found: the two articles mentioned above, and the articles by \citet{2016A&A...587A..11L}, \citet{2017A&A...603A..95A}, and \citet{2021A&A...647A.159C}. This is despite the high usage of AIA data as demonstrated by the  AIA instrument article \citep{2012SoPh..275...17L} having received over 2500 citations. \pry{We highlight here that AIA is different to the earlier \textit{EUV Imaging Telescope} \citep[EIT:][]{1995SoPh..162..291D} and \textit{Transition Region and Coronal Explorer} \citep[TRACE:][]{1999SoPh..187..229H} for which the user had a choice whether to despike the images or not as part of the software calibration process. Therefore the user could quickly assess the effect of despiking, or experiment with a variety of despiking algorithms. For AIA, the data are provided in despiked form and the user needs to actively respike the images to view the original data. The difference is partly driven by the fact that the location of the SDO orbit in the outer radiation belt leads to many more particle hits than the earlier instruments.  The large volume of AIA data thus means  it is more practical for the AIA images to be despiked prior to archiving.}

In this article we will address the following questions based on a study of a three-hour dataset from the AIA 171\,\AA\ channel:
\begin{enumerate}[label=\roman*),nolistsep]
    \item How often are solar emission features despiked?
    \item Are the features completely removed by the despiking algorithm?
    \item What types of features are despiked?
    \item Do the despiked features preferentially occur in active regions, coronal holes, or quiet Sun?
\end{enumerate}

Section~\ref{sect.data} gives technical details about the AIA instrument, including how the despiking algorithm works. Section~\ref{sect.dataset} describes the dataset used in the present article, \pry{Section~\ref{sect.stats} discusses the statistics of spikes} and Section~\ref{sect.method} gives the method for separating the true spikes from those due to energetic-particle hits. Results from the method are described in Section~\ref{sect.results}, and Section~\ref{sect.complete} discusses the completeness of the sample returned by the method. The  four questions listed above are addressed in Section~\ref{sect.prop}, where properties of the despiked solar features are presented. \pry{Section~\ref{sect.burst} gives the results from applying an EUV burst detection algorithm to the spike events, and} Section~\ref{sect.despike} describes a procedure that despikes AIA images with minimal effect on the solar features. Section~\ref{sect.summary} summarizes the findings from this article and discusses options for future work.

\section{AIA, Data Access, and Format} 
      \label{sect.data}      

AIA has four imaging telescopes that, through multiple filters, yield images in one white-light, two ultraviolet (UV), and seven extreme ultraviolet (EUV) channels. The despiking algorithm is only applied to the EUV channels. Each of these channels has a relatively narrow response function and is referred to by the wavelength in Angstroms at which the response function peaks. The 171, 193, 211, and 304\,\AA\ channels have the strongest signal in normal solar conditions, and they observe plasmas with temperatures of 0.7, 1.6, 2.0, and 0.05\,MK, respectively, based on the strongest emission lines in their passbands \citep{2010A&A...521A..21O}. From the earlier work of \citet{2013ApJ...766..127Y} and \citet{2014PASJ...66S..12Y,2014SoPh..289.3313Y} we expect that the despiking algorithm is most likely to affect compact, bright structures. The visual impression from comparing the filters' images is that compact, bright structures have a greater contrast against the background in the 171\,\AA\ channel, and so this was selected for the current article. 
Note that the 171\,\AA\ channel is dominated by the \ion{Fe}{ix} 171.1\,\AA\ emission line. The EUV image pixel sizes correspond to 0.6~arcsec or 440\,km on the Sun, and the image cadence is 12~seconds.

\textit{The Guide to SDO Data Analysis} gives details about the AIA data format and software. In particular, Section~1.3 of the \emph{Guide} describes the AIA data formats, Section~4 explains how to access AIA data, and Section~7.3 of the guide discusses the despiking and respiking of the AIA images. Some additional information on respiking  is available within the Analysis Guide for the \emph{Hinode}/\emph{EUV Imaging Spectrometer} (\urlurl{solarb.mssl.ucl.ac.uk:8080/eiswiki/Wiki.jsp?page=AIARespike}).

The AIA despiking procedure employed in the calibration pipeline is described in the \emph{Interface Region Imaging Spectrograph} (IRIS) \emph{Technical Report No.~15} \citep{iris_tr_15}. For a pixel of intensity $I$, the mean $M$ of the surrounding eight pixels is computed and a spike is flagged if $I>  (M+4)$ and $I> 1.8M$. The replacement value for the spike intensity is obtained from the 16~pixel perimeter of a $5\times 5$ box centered on the spike; the eighth lowest intensity from these 16 pixels is used. This process is repeated three times.

All AIA data used in this article were obtained from the Joint Science Operations Center (JSOC: \urlurl{jsoc.stanford.edu}), which distributes Level-1 files for the AIA images. For each EUV image, there is a corresponding spikes file that gives the location and intensity of each pixel that has been flagged as a spike. The file contains, for each spike, a one-dimensional index (within the $4096\times 4096$ image array) of the spike location; the original (Level-0) intensity of the spike; and the new (Level-1) intensity of the spike. Note that the conversion from a Level-0 to a Level-1 file does not involve any geometric corrections to the image so there is a direct one-to-one relation between the pixel locations. The  \textsf{Solarsoft} IDL routine \textsf{aia\_respike} takes the spikes file and replaces the spikes' intensities in the Level-1 file with their original values. Of course, this also restores the many actual energetic-particle hits to the image. A \textsf{Python} package for AIA data analysis \citep[\textsf{aiapy}:][]{Barnes2020} also includes software for respiking images (\urlurl{aiapy.readthedocs.io/en/latest/generated/gallery/replace\_hot\_pixels.html}).

\section{Dataset}\label{sect.dataset}

\begin{figure}[t]
    \centering
    \includegraphics[width=\textwidth]{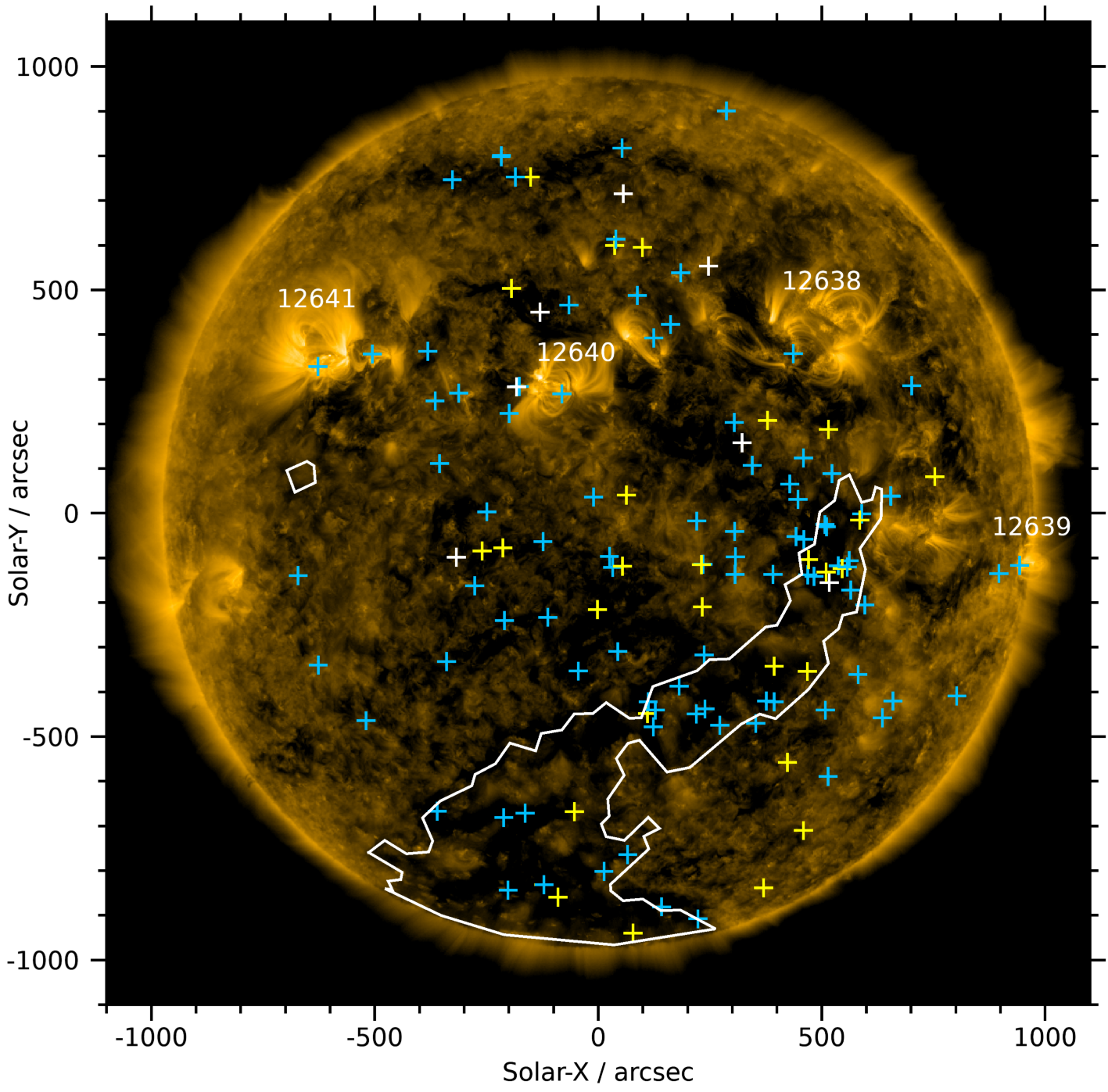}
    \caption{An AIA 171\,\AA\ image from 28 February 2017 at 10:30~UT, displayed with a logarithmic intensity scaling. The locations of two coronal holes, derived with the SPoCA algorithm, are indicated by \emph{white solid lines}. \emph{The crosses} show the locations of the 126 solar features that gave rise to the 134 three-spike events. \emph{Blue}, \emph{yellow}, and \emph{white crosses} denote event lifetimes of less than 1, 5, and 20~minutes respectively. NOAA active region numbers are shown.}
    \label{fig.fd}
\end{figure}

\begin{figure}[t]
    \centering
    \includegraphics[width=4in]{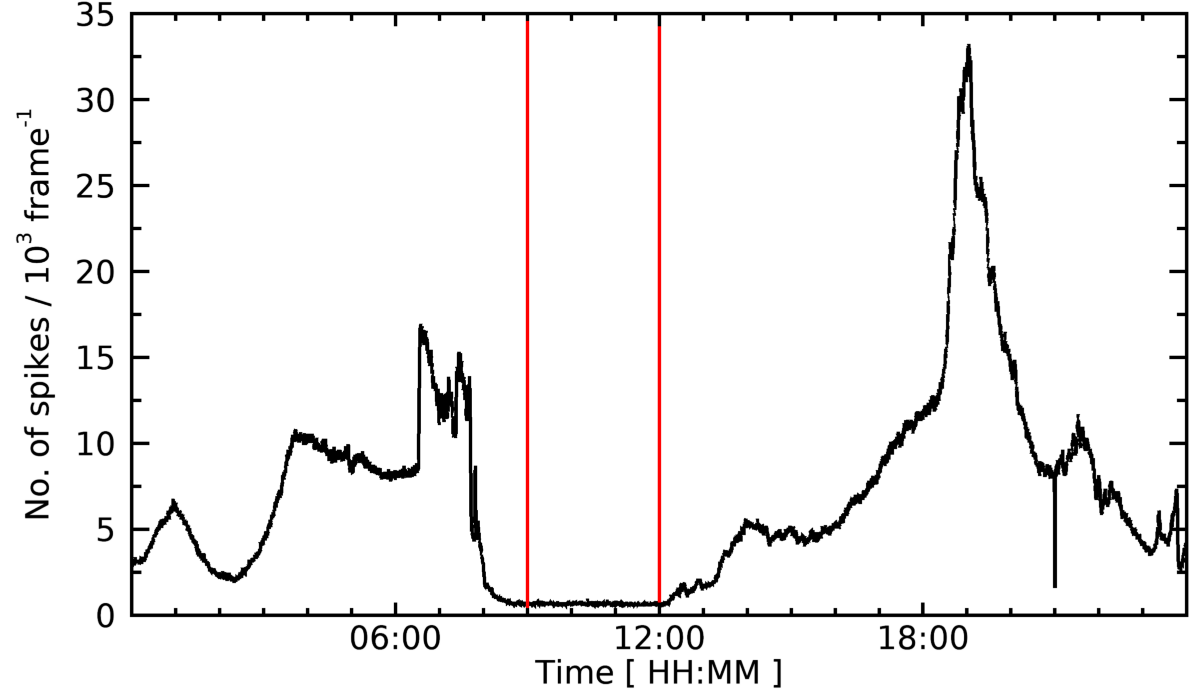}
    \caption{The variation in the number of spikes in AIA 171\,\AA\ images during 28 February 2017. The \emph{vertical red lines} show the time interval used for the event analysis. }
    \label{fig:lc}
\end{figure}

For this study we have chosen a three-hour period on  28 February 2017 when the Sun showed a  large coronal hole and several active regions (Figure~\ref{fig.fd}). The GOES 1\,--\,8\,\AA\ X-ray emission had a base level corresponding to the B1 flare class and several B-class flares occurred during the day, the largest being a B8 flare that peaked at 11:51~UT.  The coronal holes are indicated with continuous white lines and their locations were obtained through the Heliophysics Event Knowledgebase \citep{2012SoPh..275...67H}, which contains coronal-hole information from the Spatial Possibilistic Clustering Algorithm \citep[SPoCA:][]{2014A&A...561A..29V}.
There were 7191 171\,\AA\ images this day, and Figure~\ref{fig:lc} shows that the number of spikes varies by around a factor of 50 during the day.  
As SDO is in a geosynchronous orbit at a \pry{distance of 6.6 Earth radii}, then we speculate that the spikes are mostly due to energetic electrons in the outer radiation belt. The variation may be due to the latitudinal location of the spacecraft within the belt and the orientation of the spacecraft, which may modify the degree of shielding for the detector.

Intensities are given in data number (DN) units in this article. For the selected observation period, the exposure time for the 171\,\AA\ channel remained constant at 2.00~seconds.

\section{Spike Statistics}\label{sect.stats}

\pry{The observation date was chosen based on solar conditions, as described above. From the spike data shown in Figure~\ref{fig:lc} there are an average of 6807 spikes per image frame, with significant variation from 647 per frame during the period 09:00 to 12:00~UT to an average of 27,483 per frame during a 46-minute period around the peak at 19:00~UT.}

A detailed study of spike variation during the mission is beyond the scope of this article but we downloaded data for each day in 2017 February to determine if the variation on the 28th was typical for the month. In general the spike numbers were largest each day around 06:00~UT and 18:00~UT and lowest around 12:00~UT and 00:00~UT, but there was significant variation day-to-day. The periods 12\,--\,15 and 26\,--\,28 February had the lowest spike numbers, with the 28th having the sixth-lowest number. The average number of spikes ranged from 2598 (14th) to 32,463 (2nd), and the maximum number of spikes in a single frame was 311,689 (3rd), corresponding to 1.9\,\%\ of the detector pixels.

\begin{center}
\begin{table}[t]
\caption{Spike statistics for 28 February 2017.}
\begin{tabular}{c r r@{.}l r@{.}l r@{.}l}
\noalign{\hrule}
\noalign{\smallskip}
Time [UT]   & $N_{\rm avg}$ & \multicolumn{2}{c}{$P_2$} & \multicolumn{2}{c}{$N_2$} & \multicolumn{2}{c}{$P_3$} \\
\noalign{\hrule}
\noalign{\smallskip}
00:00\,--\,24:00 & 6807 & 0&937 & 2&76 & 0&00100 \\
09:00\,--\,12:00 & 647 & 0&0246 & 0&0250 & 9&62 $\times 10^{-7}$ \\
18:42\,--\,19:28 & 27,483 & 1&000 & 45&0 & 0&0711\\
\noalign{\hrule}
\end{tabular}
\label{tbl.stats}
\end{table}
\end{center}

The method for identifying true spikes in the AIA data is described in the following section and relies on spikes occurring in multiple consecutive image frames at the same pixel. We use the notation ``two-spike" and ``three-spike" to refer to spikes occurring in two and three consecutive frames, respectively.

\pry{In Table~\ref{tbl.stats} we give the probabilities for at least one two-spike [$P_2$] and at least one three-spike [$P_3$] occurring by chance when there are $N_{\rm avg}$ spikes per frame. We consider the entire day, the period 09:00\,--\,12:00~UT of low spike levels, and the peak around 19:00~UT. $P_2$ is given by $1-(1-N_{\rm avg}/4096^2)^{N_{\rm avg}}$. For $P_3$ we compute the probabilities of $i=1$, 2, 3, $\cdots$ two-spikes in a frame, multiply by $1-(1-N_{\rm avg}/4096^2)^i$, and sum over $i$. $N_2$ is the average number of two-spikes expected in each frame.}

\pry{For true spikes to be successfully identified, we want $P_n$ to be sufficiently small that random $n$-spikes are rare compared to the number of true $n$-spikes. From  Table~\ref{tbl.stats} we see that a steady rate of 6807 spikes per frame (the average value during 28 February)  would only yield seven three-spikes per day. However, the number of spikes during the 28th is highly variable, and the spike rate during the period around the peak at 19:00~UT yields at least one three-spike every 15 frames (three~minutes). To unambiguously identify true spikes during this time period would likely require four-spikes.}

\pry{For the present article we focus on the low-spike period 09:00\,--\,12:00~UT, where $P_3$ is so low that all three-spikes can be expected to be true spikes. In fact, $P_2$ is sufficiently low (one random two-spike every 40 frames) that most two-spikes will be true spikes. Despite this, we will require true spikes to be three-spikes for this article since this criterion is more suitable for typical spike levels in AIA images.}

\section{Event Identification and Analysis}\label{sect.method}

\pry{The previous section presented statistics on spike occurrence in AIA 171\,\AA\ images and we determined that all three-spikes found in our selected time interval of 09:00 to 12:00~UT will be true spikes. We thus wrote software to identify three-spikes using only the spikes data files.}

\begin{table}[t]
\caption{Processing steps for the spike analysis.}
\begin{center}
\begin{tabular}{cp{2.7in}l}
\noalign{\hrule}
\noalign{\smallskip}
\noalign{\hrule}
\noalign{\smallskip}
Step  & Description & IDL \\
\noalign{\hrule}
\noalign{\smallskip}
1 & Download spikes files for one day (JSOC) & -- \\
2 & Find three-spike spikes & \textsf{spk\_process\_sequence} \\
3 & Group spikes into events & \textsf{spk\_group\_spikes} \\
4  & Request JSOC cutouts for each event & \textsf{spk\_request\_cutouts}\\
5  & Download the cutouts & \textsf{spk\_download\_cutouts}\\
6  & Create respiked versions of the cutouts & s\textsf{pk\_respike\_cutouts}\\
7  & Create movies and light curves from the cutouts & \textsf{spk\_make\_movie\_frames}\\
\noalign{\hrule}
\end{tabular}
\end{center}
\label{tbl.steps}
\end{table}

Table~\ref{tbl.steps} summarizes the processing steps and associated IDL software routines used for the present analysis. The routines  are available through a GitHub repository \citep{2021zndo...5546641Y}, where instructions on how to use them are given. The first step is to download all of the spikes files for an AIA filter for a given day from the JSOC. These files are then processed with \textsf{spk\_process\_sequence} to identify the three-spikes. For the present article, only a three-hour set of files was processed. The check is performed by comparing with the previous and following frame. (If a spike is present for four, five, or six frames then it will lead to two, three or four three-spikes, which would be grouped together at the following step.) The method restricts the checks to heliocentric radii less than or equal to 1000~arcsec as it was found that a number of fixed hot pixels occur above the limb.

With the list of three-spikes created, the routine \textsf{spk\_group\_spikes} checks to see if the spikes are part of the same event. This is done by requiring
\begin{enumerate*}[label=\roman*)]
\item two three-spikes to occur within five arcsec of each other, and 
\item the difference in time between the end of the first three-spike and the beginning of the second is less than five minutes. These parameters were chosen to require a close physical connection between the three-spikes. We note that the parameters correspond to a velocity of 12\,km~s$^{-1}$, which is significantly greater than typical photospheric velocities of 1\,km\,s$^{-1}$ 
\pry{so multiple three-spikes from a solar feature that is being carried with the surface flow will be grouped together if they satisfy the time constraint.}
\end{enumerate*}
Section~\ref{sect.results} shows that this grouping is generally successful.

The next step is to request JSOC cutouts for each of the events, which is done by sending position and time information using the routine \textsf{sdo\_orderjsoc} (written by R.~Rutten and available from \urlurl{webspace.science.uu.nl/$\sim$rutte101/}). As the cutouts will be respiked, it is important to make sure the ``register" option is not used in the JSOC request as this performs interpolation of the images in order to smoothly track the solar rotation. The duration of the cutout sequence extends five minutes either side of the event duration. For example, a single three-spike has a duration of 24~seconds (three frames separated by 12~seconds), thus the cutout sequence will extend for 10~minutes and 24~seconds. The spatial size of the cutouts is set to $50\times 50$~arcsec$^2$.

Step~6 takes each of the downloaded cutout files, respikes them, and saves them as separate files. The respiking is performed with the \textsf{aia\_respike} routine provided by the AIA team. Movies are then created, placing the despiked and respiked frames side-by-side. Each image is centered on the first three-spike of the event's group, and solar-rotation tracking is accounted for at the pixel level by the JSOC cutout service. A box of $11 \times 11$ pixels at the center of each image frame is extracted,  averaged, and used to create a light curve for the event. This box size was chosen to give some contingency in cases where later three-spikes occurred at different locations (recall that the grouping requires three-spikes to occur within five arcsec (eight pixels) of the initial three-spike). It is also large enough to include dynamical evolution of some of the surrounding plasma emission that has not been despiked.

For ease of viewing, the images and movies are compiled into an \textsf{html} table, with rows corresponding to events. The first column gives text information about each event, the second column shows the light curve, the third column contains the side-by-side movie (with a logarithmic intensity scaling), and the fourth column shows a context image of size $200\times 200$~arcsec$^2$ centered on the event location. The latter also shows the locations of nearby spike events, indicated by their index number. All files are available on Zenodo \citep{2021zndo...5149296Y} in a file named ``spikes.zip". Upon unpacking the file, the webpage with logarithmically scaled movies is available in the directory ``20170228/0171" and the webpage with linearly scaled movies is available in the directory ``20170228\_lin/0171". The total size of the content is 194~MB.

\section{Results from Processing}\label{sect.results}

Applying the procedure described in the previous section to the three-hour sequence of 171\,\AA\ images  beginning at 09:00~UT on 28 February 2017 yielded the following results. Step~1 identified 1096 three-spikes, corresponding to 1.2~pixels per image. Step~2 grouped these three-spikes into 134 spike events and in the following sections we refer to specific spike events by indices from 1 to 134. The images and movies for each of these events are available on Zenodo \citep{2021zndo...5149296Y}.

Visual inspection of the images and movies revealed that all of the 134 spike events corresponded to dynamic features in the solar     atmosphere. None of them were data artifacts such as hot pixels or three-spike energetic-particle hits. The inspection also revealed that some events occurred close to earlier events but were not grouped either because they failed the time constraint or the spatial separation constraint. To investigate these further, we loosened the constraints applied when grouping events and found the following.
\begin{enumerate}[label=\roman*),nolistsep]

    \item Eleven events occurred within five~arcsec of an earlier event, but with a time separation greater than five~minutes.
    \item Two events occurred within five minutes of nearby events, but the spatial separation was greater than five~arcsec (but less than 15~arcsec).
    \item One event occurred within 15~arcsec of an earlier event, but separated by more than five~minutes.
\end{enumerate}
In each of these cases, we inspected movies to investigate if the event pairs corresponded to the evolution of a single solar feature. The movies were either the existing movies created by the procedure (for the Case 2 events), new movies created with \textsf{JHelioviewer} \citep{2017A&A...606A..10M}, or new JSOC cutout movies that spanned the time separation of the events (small-scale events are less easily seen in the compressed \textsf{JHelioviewer} images). The events that could be grouped are listed in Table~\ref{tbl.groups}. For example, Events 38 and 51 belonged to a quiet Sun structure that was continuously bright for the 12 minutes between them and so they clearly belong to the same solar feature. In contrast, Events 35 and 109 are coronal hole jets that, although occurring at the same location, are clearly isolated events separated by 95~minutes.

\begin{table}[t]
\caption{Additional grouping of the spike events.}
\begin{center}
\begin{tabular}{lp{2.7in}c}
\noalign{\hrule}
\noalign{\smallskip}
&&Constraint \\
Events & Comment$^{a}$ & relaxed$^{b}$ \\
\noalign{\hrule}
\noalign{\smallskip}
38, 51 & QS events separated by 12~minutes & T\\
59, 66 & Repeated brightenings within small QS structure; separated by 6~minutes & T\\
60, 70 & Part of complex AR jet (Fig.~\ref{fig.lin}); separated by 9~minutes & T\\
80, 92 & Events within a filament channel, separated by 11~minutes & T\\
81, 82, 94 & Part of evolution of the same, complex QS structure; 81 and 82 are separated by 10~arcsec, and 81 and 94 separated by 13~minutes & R, T\\
119, 134 & QS events separated by 17~minutes & T\\
129, 132 & Spatially separated by 10~arcsec but part of the same complex heating event; separated by 2~minutes & R\\
\noalign{\hrule}
\noalign{\smallskip}
\multicolumn{3}{p{4.5in}}{$^a$ Times are the difference in start times of the events; the acronyms AR, CH, and QS are used for active region, coronal hole and quiet Sun.}\\
\multicolumn{3}{l}{$^b$ T: events can occur at any time; R: events can occur within 15~arcsec of each other.}\\
\end{tabular}
\end{center}
\label{tbl.groups}
\end{table}

In total we found that 8 of the 134 spike events belonged to the evolution of a solar feature that was flagged by an earlier spike event, and these are listed in Table~\ref{tbl.groups}. Thus our total of 134 spike events corresponded to 126 unique dynamic features in the solar atmosphere.

For the remainder of this article the event indices that we use correspond to the 134 spike events and not the 126 solar features.

\section{Completeness of the Event Sample}\label{sect.complete}

The set of 134 true spike events would not be expected to be a complete sample as there are likely to be one-spikes or two-spikes that belong to dynamic solar events. In this section we investigate how many such events may have been missed.

\pry{It was noted in Section~\ref{sect.stats} that two-spikes may be a sufficient criterion to identify true spikes for the 09:00\,--\,12:00~UT period on 28 February. We modified the three-spike software to identify two-spikes and found 2159 two-spikes. The} \textsf{spk\_group\_spikes} \pry{routine then yielded 263 spike events, and these were cross-matched against the three-spike events to give 130 additional events not found by the three-spike method. We did not check how many of these are true spike events, but the probability analysis from Section~\ref{sect.stats} predicts 22 random two-spikes in the 900 image frames. This then implies 108 true two-spike events for the time interval, or 12.0 per 100 image frames. This compares with 14.9 true three-spikes per 100 image frames.}

\pry{One-spikes cannot be identified automatically using only the AIA spikes files, so it is necessary to categorize them by a visual inspection of the images. This is a time-consuming process and so the check was limited to  45 images obtained during the nine-minute period from 10:29 to 10:38~UT, which was chosen because no three-spikes occur during this time.}

Each Level-1 full-disk image was respiked and the locations of those spikes occurring within a heliocentric radius of 1000~arcsec were recorded. For each of these spikes,
images were formed in the neighborhood of the spike in both the original and respiked image. The images were then visually inspected to determine whether the spike corresponded to a dynamic event. The distinction between energetic-particle hits and true spikes was readily apparent \pry{in most cases although for some it was necessary to ``blink" consecutive images to confirm the identification}. The images used for this process, together with an IDL routine for generating the images and a text file containing the results are available through Zenodo \citep{2021zndo...5570968Y}.

\pry{The visual analysis yielded 29 one-spikes that could be grouped into 23 events (i.e. some one-spikes occurred at the locations of previous one-spikes), and seven two-spikes that could be grouped into six events. These numbers correspond to 51 one-spike events per 100 image frames and 13 two-spike events per 100 frames. The latter is in good agreement with the number derived previously, which was based on the full set of 900 image frames.
Although this is a very limited study, it shows that the number of true one-spikes is significantly larger than the numbers of true two- and three-spikes. If we take the values of 51, 12, and 15 one-, two-, and three-spikes per 100 frames then it implies that despiked events found through the three-spike method represent 20\,\%\ of the total sample.}

\section{Properties of the Spike Events}\label{sect.prop}

The Introduction posed four questions in relation to the spike events, and we address each of these in the sections below.

\subsection{Frequency}\label{sect.group}

Section~\ref{sect.results} summarized the results from the automatic processing of the AIA spikes files, and Section~\ref{sect.complete} discussed the completeness of sample identified by the three-spike method.

There were 126 unique dynamic features in the solar atmosphere that were incorrectly despiked by the AIA despiking algorithm and then identified by the three-spike method. If we assume that the method only identifies 20\,\%\ of all of the despiked solar features then we estimate that for any ten-minute period of AIA 171\,\AA\ images,  35 compact events occurring during this time will suffer from some degree of despiking. Seven of these will be identified by the three-spike method.

\begin{figure}[t]
    \centering
    \includegraphics[width=\textwidth]{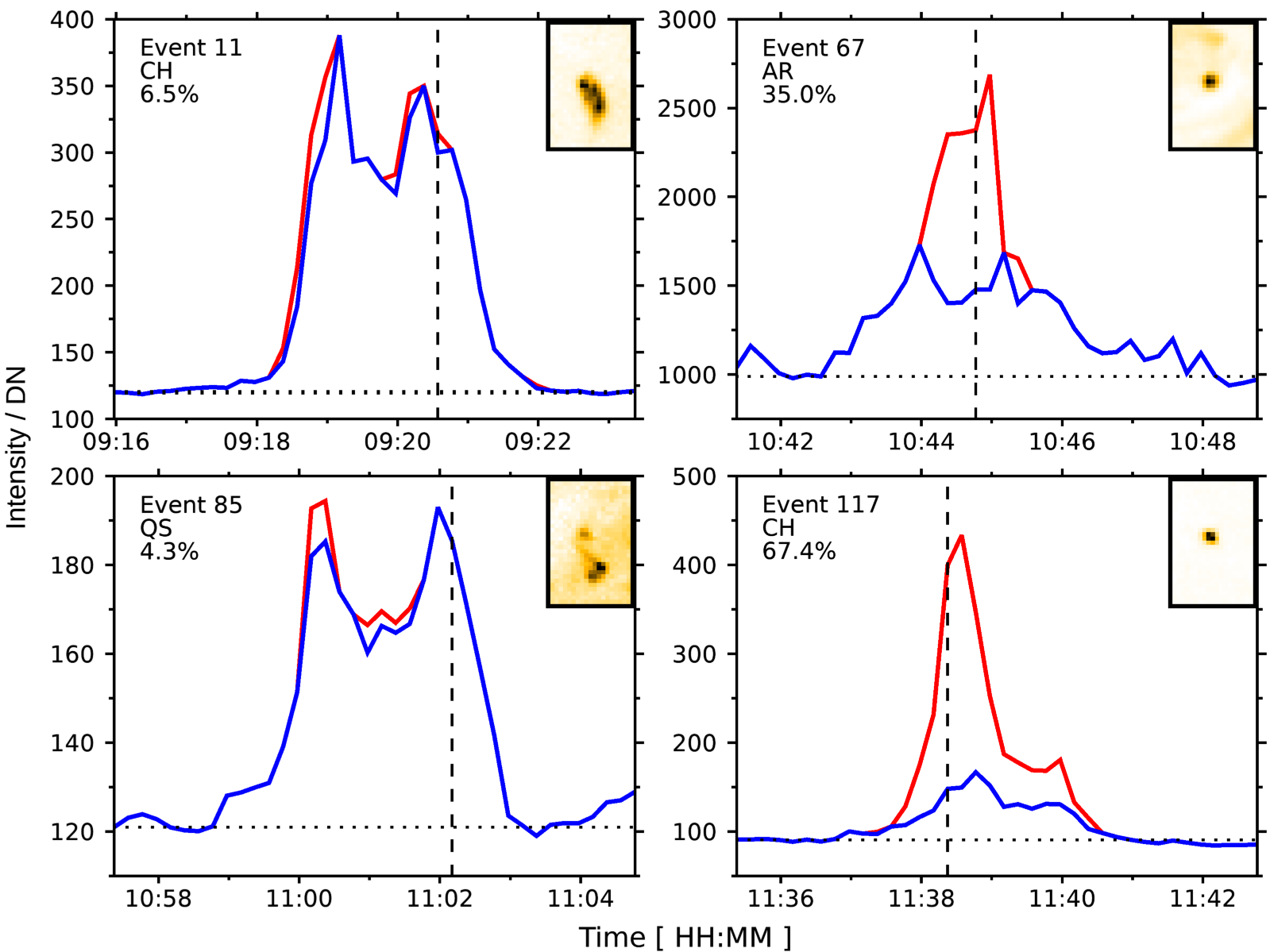}
    \caption{AIA 171\,\AA\ light curves for four of the spike events. The \emph{blue and red lines} show the light curves for the despiked and respiked image sequences, respectively. The \emph{horizontal dotted line} shows the background level used in computing the integrated event intensities. The \emph{small inset images} show the events at the times indicated by the \emph{vertical dashed lines}. Each image is $11\times 17$~arcsec$^2$ and is displayed with a reversed-logarithmic scaling.}
    \label{fig.lc}
\end{figure}

\subsection{Visibility and Intensity Depletion}\label{sect.vis}

Inspection of the light curves and movies available on Zenodo \citep{2021zndo...5149296Y} shows that the despiking routine does not completely remove compact brightenings from the images. Even the smallest brightenings in this three-hour dataset can be identified in both the despiked and respiked image sequences since only a fraction of the image frames are affected. The despiked frames  typically show compact brightenings with a ``cratered" appearance where the central pixel or pixels have a lower intensity than the immediately surrounding pixels. 

\begin{figure}[b]
    \centering
    \includegraphics[width=4in]{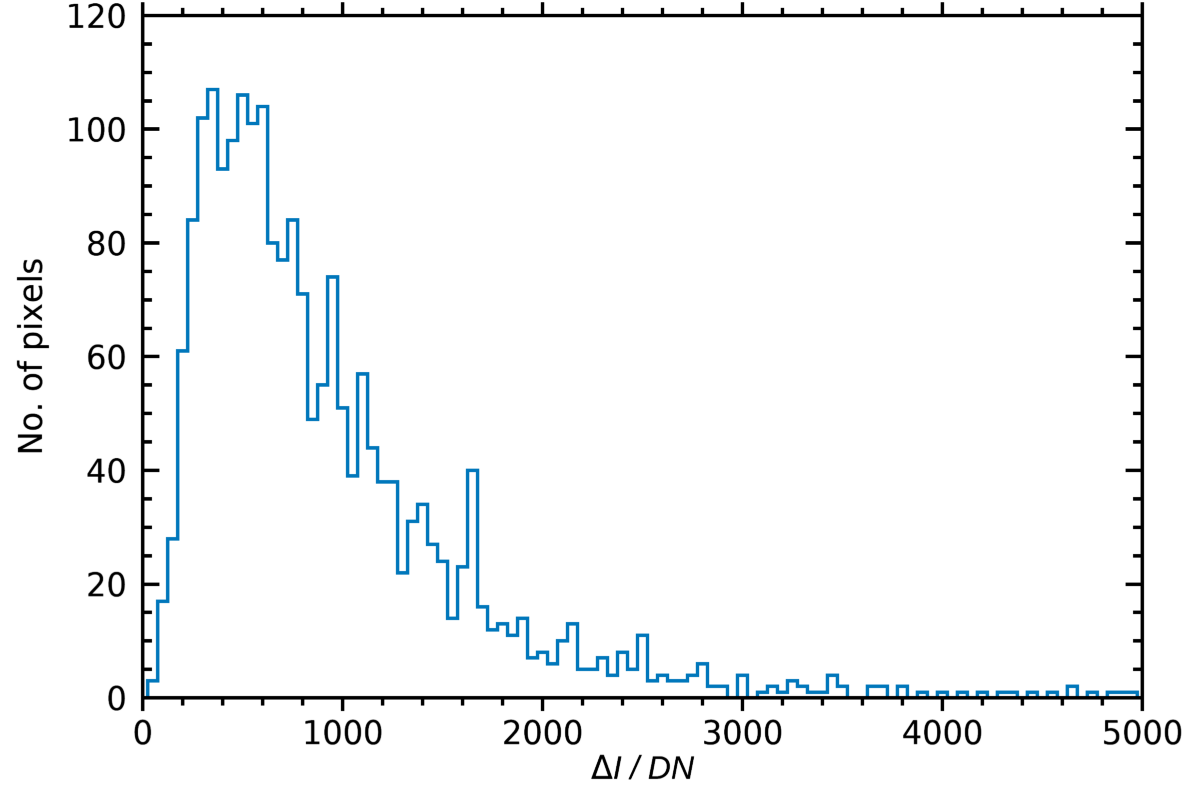}
    \caption{The distribution of intensity differences [$\Delta I$] between the true intensities and the despiked intensities. The distribution tail extends to the highest value of 12,374~DN (not shown), and 98.7\,\%\ of the total pixels are displayed.}
    \label{fig.dist}
\end{figure}

The light curves for the approximately ten minutes around the despiking period are often complex, with only 15 events showing simple rise-and-fall behavior against a uniform background intensity. This reflects the fact that the intense, compact brightening that is typically despiked is often accompanied by associated loop, jet or cloud-like emission that evolves alongside the brightening. Even for the morphologically simple events, there may be structure below the resolution of the instrument that gives rise to complex light curves. Four of the simpler light curves are shown in 
Figure~\ref{fig.lc}: two from the coronal hole, one from quiet Sun, and one from an active region. In each case the event brightens and fades in less than five minutes.
The curves differ slightly from those provided on Zenodo \citep{2021zndo...5149296Y} as the pixel boxes used to define the events \pry{were in this case customized to enclose the spatial extents of the events}. The blue lines show the light curves derived from the despiked data and the red lines are for the respiked data. That is, the red line shows the true intensity evolution of the events. The percentage differences in the intensities integrated over the events' lifetimes are displayed on each panel. Event 117 was the most impacted of all the 134 spike events, with the intensity reduced by 67\,\%\ over the four-minute lifespan. Despite this large reduction in intensity, the event remains easily discernible in the despiked movie \citep{2021zndo...5149296Y}, demonstrating that the despiking did not make the feature invisible \pry{although, of course, the missing intensity would have a significant impact on differential emission measure analysis, for example}. Event 11 from Figure~\ref{fig.lc} was a small jet in the coronal hole; Event 67 was the brightest of all of the spike events (see below); and Event~85 was a compact quiet Sun loop.

The intensity removed from each pixel by the despiking algorithm varies \pry{from 77~DN to over 12,000~DN}. For each of the 1096 pixels identified by the three-spike method there are three intensity measurements, giving 3288 measurements in all. However the number of unique pixels is 2126 as three-spikes can overlap in the case that four or more consecutive frames have the same detector pixel despiked.
Figure~\ref{fig.dist} shows the distribution of the intensity difference [$\Delta I$] between the original pixel intensity and the despiked pixel intensity for these 2126 pixels.
The high-intensity tail of the distribution extends beyond the plot boundary to a maximum value of 12,374~DN,  which belonged to Event~67 (see also Figures~\ref{fig.lc} and \ref{fig.lin}). For reference, we note that the AIA CCDs saturate just below 16,383~DN. This event had 18 of the greatest 19 $\Delta I$-values, all of which were larger than 6000~DN. (Despite these large differences, the integrated intensity over the event's lifetime was reduced by only 35\,\%, as shown in Figure~\ref{fig.lc}.)

The median of the $\Delta I$-distribution is 752~DN, and the smallest value is 77~DN. The latter occurred for  Event~102, which occurred near the boundary of the coronal hole.  It had 10 of the 12 smallest $\Delta I$ values.

\begin{table}[t]
\caption{A comparison of the events with the largest and smallest total intensity differences caused by despiking.}
\begin{center}
\begin{tabular}{cccrlcccr}
\noalign{\hrule}
\noalign{\smallskip}
\multicolumn{4}{c}{Largest events} & &\multicolumn{4}{c}{Smallest events} \\
\cline{1-4}\cline{6-9}
\noalign{\smallskip}
Event & Region$^a$ & $n_{\rm pix}$& $\Delta I_{\rm tot}$ [DN] & & Event & Region$^a$ &$n_{\rm pix}$& $\Delta I_{\rm tot}$ [DN] \\
\noalign{\hrule}
\noalign{\smallskip}
 67 & AR &  34 &212,886 & & 121 & CH &3& 578 \\
130 & QS & 121 &139,077 & &  10 & CH &3& 581 \\
124 & QS &  61 &124,290 & &  53 & CH &3& 854 \\
119 & QS &  57 & 90,098 & &  80 & QS &3& 948 \\
117 & CH &  96 & 87,899 & &  44 & QS &3&1009 \\
\noalign{\hrule}
\noalign{\smallskip}
\multicolumn{9}{l}{$^a$ AR -- active region; QS -- quiet Sun; CH -- coronal hole.}
\end{tabular}
\end{center}
\label{tbl.totint}
\end{table}

We define $\Delta I_{\rm tot}$ to be the total intensity lost to three-spikes during an event's evolution, a quantity that can be computed directly from the AIA spikes files without the need to access the AIA image files. For example, if an event has a single three-spike during its lifetime, then $\Delta I_{\rm tot}$ is the total intensity lost in the three despiked pixels. Care was taken to avoid double-counting pixels from consecutive three-spikes in the same spatial pixel. For example, if there are two consecutive three-spikes during the evolution then this means that there are four consecutive despiked pixels (not six) and so only these four values enter into the $\Delta I_{\rm tot}$ calculation. 
The events with the largest and smallest  $\Delta I_{\rm tot}$ values are listed in Table~\ref{tbl.totint}.
As expected, given the discussion above, active-region Event~67 has the largest intensity decrease. Interestingly, though, the next four most-affected events occurred in quiet Sun or coronal hole regions. (Event 117 is the event shown in Figure~\ref{fig.lc}, and it had the largest percentage intensity decrease of all the events.) Table~\ref{tbl.totint} also gives the number of despiked pixels [$n_{\rm pix}$] \pry{flagged by the three-spike method for each event}.  It can be seen that this is \pry{an important factor in determining the size of the intensity decrease:}
\pry{the weakest events all have just a single three-spike [$n_{\rm pix}=3$], while the quiet-Sun and coronal-hole large-intensity events have between 57 and 121 despiked pixels.}

\subsection{Event Types}\label{sect.types}

\begin{figure}[t]
    \centering
    \includegraphics[width=\textwidth]{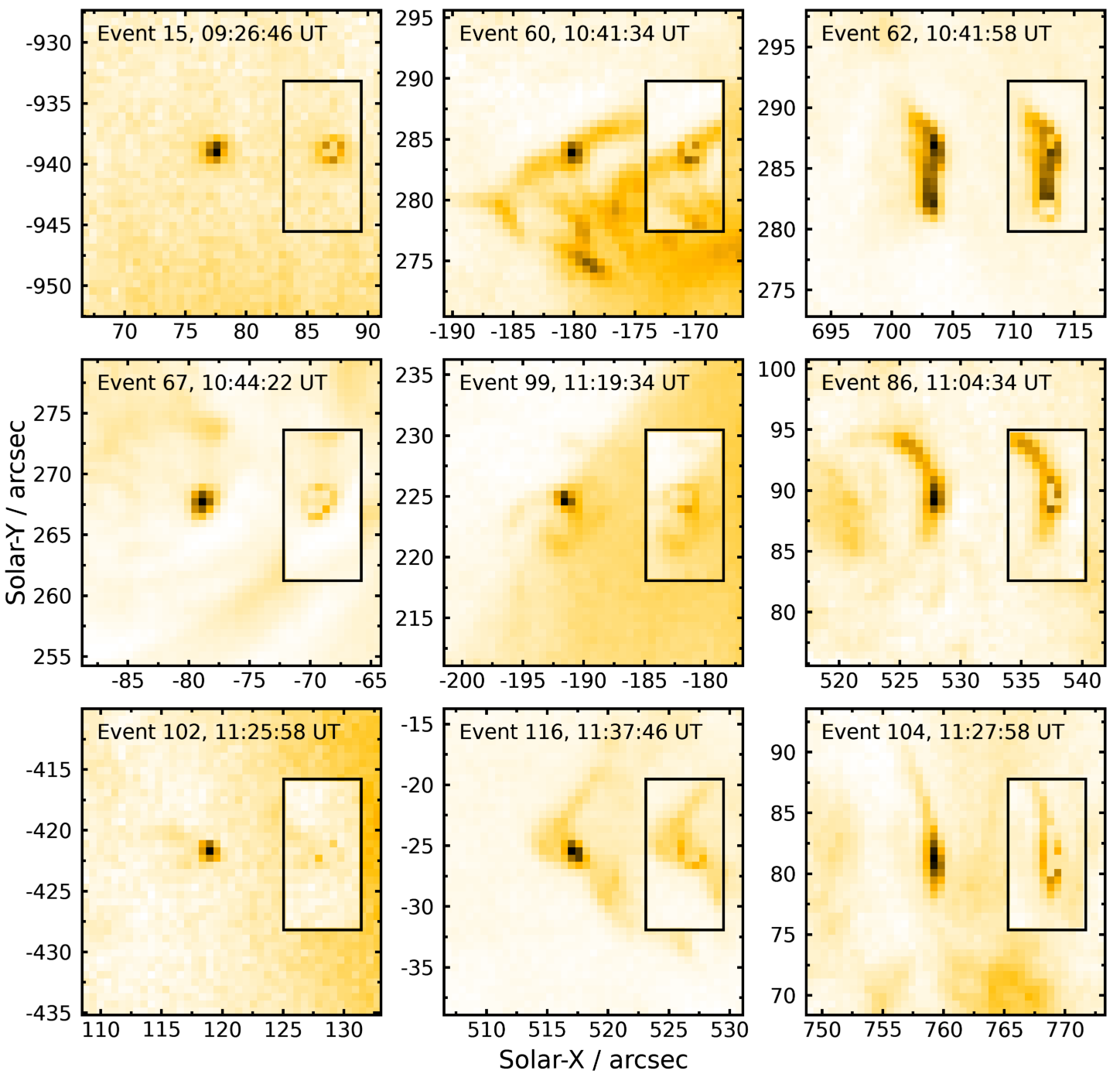}
    \caption{Nine examples of respiked solar features. Images are shown with a reversed, linear intensity scaling. Each panel shows the respiked image and the dark structures to the left of center are the features that were found by the three-spike method. The inset images to the right of center show how the structures look in the despiked images.
    The left and middle columns show examples of point-like sources, and the right column shows examples of spatially extended sources.}
    \label{fig.lin}
\end{figure}

Movies with a linear intensity scaling \citep{2021zndo...5149296Y} were inspected in order to investigate the morphology of the structure from which the pixels had been despiked. All but six of the 134 events (95.5\,\%) showed a point-like structure of size two arcsec or less in size, and examples are shown in the left and middle columns of Figure~\ref{fig.lin}. The remaining events had an elongated structure, and three examples are shown in the right column of  Figure~\ref{fig.lin}. In each of the nine panels the solar feature that had been despiked is shown, in its respiked form, to the left of center of the image. The small inset images show the despiked images of the features, revealing the significant impact on the features' intensities, but also it can be seen there is usually residual signal that allows the features' locations to be identified.

\begin{figure}[t]
    \centering
    \includegraphics[width=\textwidth]{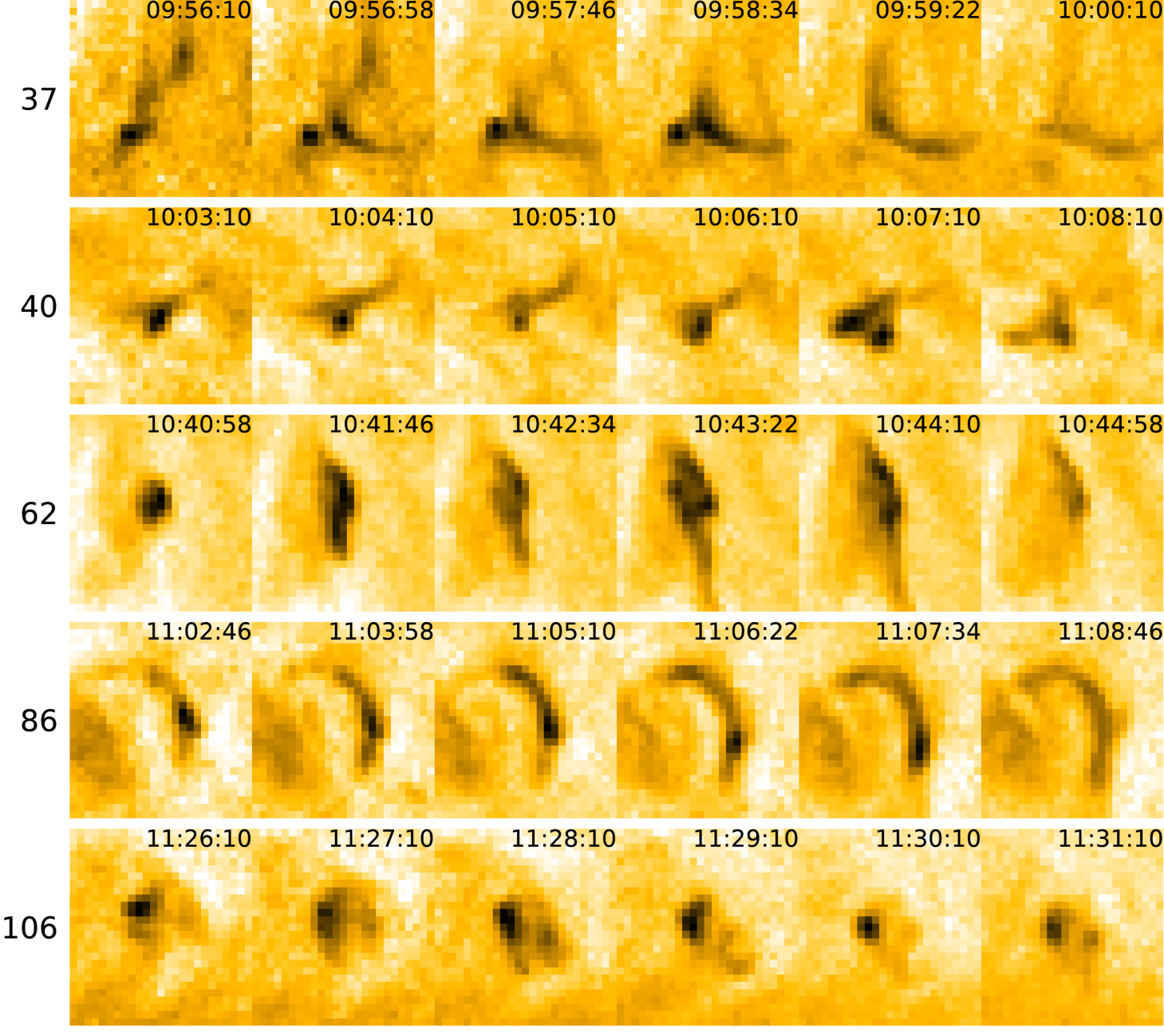}
    \caption{AIA 171\,\AA\ images from the evolution of five of the spike events. The images are arranged as rows for each event, with the event number given on the left. The observation times, in UT, are given in the top-right corner of each frame, and  reversed, logarithmic intensity scaling is used. The image size is $14.4\times 15.6$~arcsec$^2$ in each case.}
    \label{fig.evol}
\end{figure}

The morphology referred to here belongs to the brightening that has been wrongly despiked by the AIA algorithm. In most cases this brightening is part of the evolution of a larger structure that may include loops, jets, or cloud-like emission. These structures are usually fainter and are best seen in the movies with a logarithmic intensity scaling \citep[available from][]{2021zndo...5149296Y}. Figure~\ref{fig.evol} shows six image frames from five different spike events to give an indication of the diverse morphology of the solar features that can be despiked.

One type of event that we highlight are coronal-hole jets \citep[see the review of][]{2016SSRv..201....1R}. These exhibit a short-lived, dynamic ``spire" of emission that typically arises from a coronal bright point. The latter often exhibits one or more intense, compact brightenings during the eruption, and \citet{2014PASJ...66S..12Y,2014SoPh..289.3313Y} found that these brightenings were partially despiked in the AIA images of the two jets that they studied.

We found seven spike events that corresponded with coronal hole jets: numbers 11, 35, 64, 73, 109, 111, and 116. 
Events 11 and 73 were small jets with bright bases and short durations of five~minutes. Events 35 and 109 occurred at the same location  but separated by 95~minutes. Although the jets' bases are outside of the SPoCA coronal hole, their morphology is consistent with coronal-hole jets and we believe that line-of-sight plage emission just to the west of the northernmost extension of the coronal hole likely prevented the SPoCA algorithm from identifying the region. Event 116 gave rise to the largest jet, and the despiked pixels were in one location of a large bright point about 20~arcsec across (Figure~\ref{fig.lin}). Events 64 and 111 were weaker jet events that also showed single, intense brightenings on one side of the jet base. 

\subsection{Spatial Locations}\label{sect.loc}

In Section~\ref{sect.results} we found that the 134 grouped three-spike events corresponded to 126 unique solar features. Figure~\ref{fig.fd} shows the locations of these features on the solar disk (each has been adjusted for solar rotation to match the displayed image).
Coronal holes flagged by the SPoCA software \citep{2014A&A...561A..29V} and obtained through the Heliophysics Event Knowledgebase \citep{2012SoPh..275...67H} IDL software are shown as single white contours. The south polar coronal hole had a large extension reaching to the equator, and the other SPoCA coronal hole to the south of AR 12641 was very small (no spike events are found near it). 

\pry{For the three basic types of solar region -- active region, quiet Sun, and coronal hole -- the breakdown of the 126 solar features is 6\,\%, 59\,\%, and 35\,\%, respectively.}
 The coronal hole events were initially identified from Figure~\ref{fig.fd} based on their locations relative to the SPoCA coronal hole. The images and movies provided by \citet{2021zndo...5149296Y} were then inspected to confirm or reject the initial identification. We found 44 coronal hole events, which included 14 at or close to the coronal hole boundary. The events can be spatially divided into those with solar-$y$ positions below $-600$~arcsec (12 events), between $-500$ and $-300$~arcsec (12 events) and between $-300$ and $+100$~arcsec (20 events). The large number of events in the latter, relatively compact spatial region is striking, and it suggests that there is an unusually high degree of dynamic solar activity occurring here.

Only eight events (6\,\%\ of total) were associated with the four numbered active regions in Figure~\ref{fig.fd}.  AR 12641 had two events: one in the core (Event~78) and another in a peripheral region to the west (Event~96). A B8 flare occurred in AR 12641, peaking at 11:51~UT, but no spike events are associated with it.  AR 12640 had three events, one of which was the brightest of all the respiked events (Event~67; Figure~\ref{fig.lin}). The other two events (Events~58 and 60) were both part of the evolution of a complex jet extending to the east of the AR core and which was visible for several hours from 10:00~UT. Figure~\ref{fig.lin} shows an image from Event 60. Event~99 occurred  at the boundary of bright loops extending southwards from  AR 12640. As this event is highly structured (unlike the smooth intensity distribution of the loops) and it partly extends beyond the sharp intensity edge of the loops (Figure~\ref{fig.lin}),  we consider it to be a low-lying, quiet Sun event that shares the same line-of-sight as the loops.

A small, unnumbered AR to the northwest of 12640 produced Event 128, which was just a simple point-like brightening. Event 29 occurred in the plage of decayed AR 12638 and again was a simple point-like brightening. The final AR event was Event~120, which occurred close to the limb in the bright plage of AR 12639. 

The remaining 74 solar features are nominally identified with quiet Sun. Three of these  appear to be connected with filament channels: one extending north from AR 12640 (Events 80 and 107), and the other extending to the east of the coronal hole around $y=-300$ (Event 37).

\section{Comparison with EUV Bursts}\label{sect.burst}

\pry{
EUV bursts were identified from AIA 171\,\AA\ images by \citet{2021A&A...647A.159C}  by searching for individual spatial pixels that brighten above the background intensity level by a specified amount during a 30-minute interval. Neighboring pixels with a similar time behavior were grouped, giving burst sizes ranging from 0.2~Mm$^2$ (1 pixel) to 10~Mm$^2$ (53 pixels). Lifetimes ranged from 36 to 400~seconds, with a median of 120~seconds.}

\pry{The algorithm was applied to the spike events here to investigate how many  are EUV bursts.
Since \citet{2021A&A...647A.159C} used  30-minute sequences of images for the burst detection, here we created 30-minute duration cutout movies for each event, centered on the time of the first three-spike in the events' sequences. This is important for burst detection as the algorithm requires access to background intensity levels prior to the burst appearance, and the 10-minute image sequences produced by the spike IDL routines were often not sufficient to yield a good measure of the background.
}

\pry{
The burst algorithm yields a movie for each event with the burst locations identified by blue contours. The output from the spike IDL routines was used to overlay the locations of three-spikes as crosses. Immediately apparent from the movies is that bursts are much more numerous than the spike events, with typically 10\,--\,50  bursts in the 30-minute image sequences over the $50\times 50$~arcsec$^2$ fields-of-view. Inspection of the  movies showed that 128 of the 134 spike events corresponded with bursts. In some cases the burst was not exactly at the same location as a three-spike, but it occurred at the same time and within 2~arcsec \pry{(the typical size of the spike events)}. 
}

\pry{
Of the remaining six spike events, the three associated with the complex active region jet (Events 58, 60, and 70) did not have bursts at the same location, although there were many bursts occurring nearby during the movie sequences. Events 80, 105, and 126 all occurred in the quiet Sun. Event~80 was the second weakest spike event (Table~\ref{tbl.totint}) and thus likely failed to reach the threshold intensity for burst detection. Events 105 and 126 both featured small bursts at the same time as the three-spikes, but around 3\,--\,5~arcsec further away and so we consider them distinct features.
}

\pry{
In summary, 96\,\%\ of the spike events can be classified as EUV bursts according to the criteria of \citet{2021A&A...647A.159C}.
}

\section{Despiking Respiked Data}\label{sect.despike}

The AIA despiking algorithm is hard-coded into the processing pipeline that generates the Level-1 data and it is unlikely to be modified as a change would then require the entire archive of AIA EUV images to be reprocessed. While respiking restores signal to despiked solar events, it also restores the energetic-particle hits. For aesthetic reasons, when creating movies from respiked data it is preferable to remove the energetic-particle spikes while retaining the true data spikes, i.e. apply a despiking routine that does a better job of cleaning the images than the AIA pipeline algorithm.

We highlight here a routine in the \textsf{Solarsoft} IDL distribution called \textsf{aia\_clean\_ cutout\_sequence} that is specifically for cleaning cutout sequences that have been downloaded from the JSOC and then respiked with the \textsf{aia\_respike} routine. The crucial difference with the pipeline algorithm is that it compares a pixel's intensity with the previous and following image frames, which is usually sufficient to avoid despiking solar atmospheric features. Otherwise the method is very similar to the AIA team's algorithm. The particular steps and parameters used are:
\begin{itemize}
    \item The intensity in a pixel must be greater than that of the same pixel in both the preceding and subsequent exposures.
    \item The intensity in a pixel must be greater than the average intensity of the preceding and subsequent exposures by at least a factor of 1.7. (For the first and last frames in the sequence this factor is raised to 2.5.)
    \item The intensity in a pixel must be greater than the median of the surrounding 5$\times$5 pixel block by 5 DN.
\end{itemize}

An alternative technique for removing spikes from respiked AIA images, also making use of time series data, was described by \citet{2021A&A...647A.159C}. \pry{\textsf{Solarsoft} contains the routine} \textsf{trace\_unspike\_time} \pry{that was developed for the TRACE mission and is described in \citet{2000ApJ...535.1027A}. This also acts on time series of images and we tested it on the spike event movies. We found that 125 of the 134 events were partially despiked by this routine (as determined from the light curves, as described above) and so it was less successful than} \textsf{aia\_clean\_cutout\_sequence}. \pry{However, the light curve comparison demonstrated that the routine was a significant improvement on the  AIA despiking algorithm. Since} \textsf{trace\_unspike\_time}  \pry{was developed for TRACE, then the algorithm parameters may need to be adjusted for AIA data.}

\section{Summary and Discussion}\label{sect.summary}

\pry{EUV images from AIA are despiked prior to archiving in order to remove energetic-particle hits on the detector (spikes). Sometimes solar features are despiked too, modifying their intensity and morphology. The present article is the first systematic study of an AIA dataset to quantify how many solar features are affected and how significant the despiking is. By giving statistics and examples, including a complete set of movies and light curves available on Zenodo \citep{2021zndo...5149296Y}, we hope the results will give other researchers guidance on whether their datasets would benefit from respiking. In addition, our software procedures for finding solar events that have been wrongly despiked \citep{2021zndo...5546641Y} may enable new statistical studies.}

\pry{Despiked solar events were identified by finding  detector pixels that were despiked in three consecutive images (three-spikes).}
From a three-hour sequence of 171\,\AA\ images from 28 February 2017,  1096 such three-spikes were found. These were then grouped into 134 spike events by requiring three-spikes to occur within five~arcsec and five~minutes of each other. Visual inspection of image sequences showed that these spike events belonged to 126 unique solar events.

The article Introduction asked four questions of the despiked solar events. Based on the time period analyzed here and, assuming it is representative of AIA data in general, the answers are: 
\begin{enumerate}[label=\roman*),nolistsep]
    \item \emph{How often are solar emission features despiked?} On average, in a ten-minute period of AIA 171\,\AA\ images, 35 dynamic, compact features are despiked. The three-spike method identifies 20\,\%\ of these.
    \item \emph{Are the features completely removed by the despiking algorithm?} No, the features are always identifiable as they are only despiked for a fraction of their lifetimes, and there are always residual pixels next to the despiked pixels. The intensity reduction over a despiked event's lifetime in this study was up to 67\,\%, although it was usually much smaller.
    \item \emph{What types of features are despiked?} 96\,\%\ of the features despiked are tiny brightenings about two arcsec or less in size. The remaining features are elongated in one direction. \pry{EUV bursts comprise 96\,\%\ of the events, and none were related to flares.} The despiked feature is usually one aspect of a larger structure that may include loops, jets, or cloud-like emission.
    \item \emph{Do the despiked features preferentially occur in active regions, coronal holes or quiet Sun?} Events are found in all three locations, with percentages 6\,\%, 35\,\%, and 59\,\%, respectively. In relation to solar-surface area, there is a preference for the events to occur in the coronal hole, particularly the narrow, equator-ward extension in the present dataset.
\end{enumerate}

\pry{In Section~\ref{sect.burst} we used the EUV burst detection algorithm of \citet{2021A&A...647A.159C} to find that 96\,\%\ of the spike events coincided, spatially and temporally, with EUV bursts. The latter can be considered generic small-scale, short-lived  atmospheric brightenings and so the high correlation  shows there is nothing unusual about the set of spike events. }

\pry{There are many other methods that have been used to identify transient brightenings in the solar atmosphere, and we highlight here the recent article of \citet{2021arXiv210403382B}. They described ``campfires" identified from images obtained with} the \textit{Solar Orbiter Extreme Ultraviolet Imager} (EUI). The data set had a 245~second duration, a cadence of five seconds and a pixel size of 200\,km (compared to a cadence of 12~seconds and pixel size of 435\,km for AIA), and events were detected with lifetimes of 10 to 200~seconds, and sizes of 400 to 4000\,km. The EUI bandpass is centered at 174\,\AA, giving a greater contribution from \ion{Fe}{x} lines at 174.5 and 177.2\,\AA\ than the AIA 171\,\AA\ filter, and thus access to slightly hotter plasma. The campfires were detected by spatial intensity enhancements in individual image frames rather than using time series \pry{and the relation with EUV bursts has not been established, although they are likely very similar.}

\pry{The spike event detection method is quite distinct from the EUV burst and campfire methods since the user does not specify an event type through intensity thresholds -- the criteria are imposed by the AIA despiking algorithm with no consideration of solar feature types.}
The despiked events do not offer a complete sample of bursts or campfires, but the capability of rapidly identifying a large number of events through a novel technique unbiased by human selection effects \pry{and a relatively compact data set} could yield unique insights into the occurrence of small-scale events through the solar cycle. \pry{By studying each individual event within a prescribed time period, we have verified in this article that the three-spike method can efficiently isolate solar events using only the information in the AIA spikes files.}

This initial survey has only used the AIA 171\,\AA\ filter data to identify the spike events. Further work will take into account the other AIA filters, with  193 and 211\,\AA\ likely the most useful due to a stronger signal. For example, is there a one-to-one relation between 171\,\AA\ three-spikes and 193\,\AA\ three-spikes, or does the latter channel reveal a distinct group of events?
It also may be possible to identify one- or two-spike events using only the AIA spikes files if they are seen in two or more filters, which is not possible with a single filter, as discussed in Section~\ref{sect.method}.
Due to spatial offsets between the filters, however, additional processing will be required to group spikes based on physical location, not just detector location.

A long-term goal is to create and maintain a database of true data spikes (such as identified through the three-spike method) for each AIA EUV channel that stores their times and spatial locations. This is likely beyond the capability of a standard desktop computer using IDL code, and the present authors are investigating filtering and clustering procedures running on graphics processing units (GPUs). This spike database could be automatically matched to information in the HEK in order to assign the spike events to solar features, such as active regions, coronal holes and flares. \pry{As touched upon in Section~\ref{sect.stats}, there are periods of high spike numbers when three-spikes will not be sufficient to distinguish true spikes from random spikes. The procedure in this case is to filter out these periods from the analysis. This is a trivial step, however, as the number of spikes is recorded in the spike data files.}

The analysis performed here made use of JSOC cutout images and potentially a machine learning procedure could be applied to classify the events' morphologies into different categories. If we extrapolate the 134 events found here for a three-hour time sequence in one EUV channel to the entire 11-year mission and seven EUV channels, then we have 30~million events. It is simply not feasible to request and respike so many datasets, so such analyzes would be restricted to much shorter time periods.

\pry{Finally, we highlight that the three-spike method presented here is a means to mining the AIA spikes files for real solar events that may then be used for some type of statistical survey. If a user has a particular data set and they have concerns that some of the data have been incorrectly despiked, then the procedure is to first respike the data (with} \textsf{aia\_respike}\pry{) and then despike with} \textsf{aia\_clean\_cutout\_sequence} \pry{(or an alternative routine). As shown in Section~\ref{sect.despike}, this procedure largely results in the true solar features remaining untouched by the despiking process.}


\begin{ack}
The authors thank Barbara J.\ Thompson for valuable discussions, and  We thank the reviewer for valuable comments and suggestions. We also thank Paul Boerner and Mark Cheung of the AIA instrument team.
\end{ack}

\begin{fundinginformation}
 We acknowledge support from  NASA's Heliophysics Guest Investigators program and the  GSFC Internal Scientist Funding Model competitive work package program. E.I.~Mason’s research was supported by an appointment to the NASA Postdoctoral Program at the NASA Goddard Space Flight Center, administered by the Universities Space Research Association under contract with NASA.
\pry{This project has received funding from the European Research Council (ERC) under the European Union’s Horizon 2020 research and innovation programme (grant agreement No 695075).}
\end{fundinginformation}

\begin{dataavailability}
The AIA images and spikes files used  in this article are publicly available from the JSOC at \urlurl{jsoc.stanford.edu}. Processed datasets generated during this project are available at \urlurl{zenodo.org} \citep{2021zndo...5570968Y,2021zndo...5149296Y}.
\end{dataavailability}

\begin{ethics}
\begin{conflict}
The authors declare that they have no conflicts of interest.
\end{conflict}
\end{ethics}

\bibliographystyle{spr-mp-sola}
\bibliography{sola_bibliography_example}

\end{article} 

\end{document}